# Impact of climate change on West Nile virus distribution in South America

Camila Lorenz[1]*, Thiago Salomão de Azevedo[23]*, Francisco Chiaravalloti-Neto[1]

1 Departamento de Epidemiologia, Faculdade de Saúde Pública, Universidade de São Paulo, Av. Dr. Arnaldo, 715, São Paulo CEP 05509-300, Brazil.

2 Secretary of Health, Municipality of Santa Barbara d'Oeste - CEP 13450-021, Sao Paulo, Brazil.

3 Laboratório em Entomologia e Sistemática Molecular, Faculdade de Saúde Pública, Universidade de São Paulo, Av. Dr. Arnaldo, 715, São Paulo CEP 05509-300, Brazil.

*These authors contributed equally.

*Corresponding author: Camila Lorenz. Departamento de Epidemiologia, Faculdade de Saúde Pública, Universidade de São Paulo, Av. Dr. Arnaldo, 715, São Paulo CEP 05509-300, Brazil. Phone: +551130617913 E-mail: camilalorenz@usp.br.

## ABSTRACT

West Nile virus (WNV) is a vector-borne pathogen of global relevance and is currently the most widely distributed flavivirus of encephalitis worldwide. This virus infects birds, humans, horses, and other mammals, and its transmission cycle occurs in urban and rural areas. Climate conditions have direct and indirect impacts on vector abundance and virus dynamics within the mosquito. The significance of environmental variables as drivers in WNV epidemiology is increasing under the current climate change scenario. In this study, we used a machine learning algorithm to model WNV distributions in South America. Our model evaluated eight environmental variables (type of biome, annual temperature, seasonality of temperature, daytime temperature variation, thermal amplitude, seasonality of precipitation, annual rainfall, and elevation) for their contribution to the occurrence of WNV since its introduction in South America (2004). Our results showed that environmental variables can directly alter the occurrence of WNV, with lower precipitation and higher temperatures associated with increased virus incidence. High-risk areas may be modified in the coming years, becoming more evident with high greenhouse gas emission levels. Countries such as Bolivia and Paraguay will be greatly affected, drastically changing their current WNV distribution. Several Brazilian areas will also increase the likelihood of presenting WNV, mainly in the Northeast and Midwest regions and the Pantanal biome. The Galápagos Islands will also probably increase their geographic range suitable for WNV occurrence. It is necessary to develop preventive policies to minimize potential WNV infection in humans and enhance active epidemiological surveillance in birds, humans, and other mammals before it becomes a more significant public health problem in South America.

**Abbreviations:** IPCC, Intergovernmental Panel on Climate Change; GCM, global climate model; GGE, greenhouse gas emission; MIROC-5, Model for Interdisciplinary Research On Climate; PCA, principal component analysis; RCP, Representative Concentration Pathway; WNV, West Nile virus

## 1. Introduction

Currently, West Nile virus (WNV) is distributed throughout all continents, excluding Antarctica, and is considered the most widely spread arbovirus on the planet (Kramer et al., 2008). Its related disease, West Nile fever, is commonly asymptomatic or mild, but approximately 1 in 150 patients diagnosed with West Nile fever becomes neuroinvasive, inducing encephalitis or even death (CDC, 2018). Although mosquito bites are responsible for almost all human infections, WNV transmission also occurs through organ transplantation, blood transfusion, intrauterine transmission, and possible transmission through breastfeeding (Hayes et al., 2005).

WNV was first isolated in 1937 from a febrile woman in Uganda (Smithburn et al., 1940) and was subsequently associated with sporadic cases of disease and major outbreaks in Africa, the Middle East, Eurasia, and Australia. From 1937 to 1999, WNV cumulated slight medical attention as the cause of sporadic encephalitis and febrile illness in Africa, Asia, and Europe (Hayes et al., 2005). After the unexpected detection of WNV in New York City in 1999, this flavivirus has extended rapidly westward through the United States, southward into Central America and the Caribbean, and northward into Canada, causing the largest outbreak of neuroinvasive WNV disease ever reported (Hayes et al., 2005).

Since its appearance in the American continent in 1999, the virus has resulted in over 48,000 reported cases, 24,000 reported neuroinvasive cases, over 2,300 deaths (CDC, 2018), and an estimated 7 million total human infections in the USA only (Ronca et al., 2019). At present, WNV is considered one of the most relevant viruses to cause zoonotic diseases in the US population (CDC, 2018). WNV is also an important animal pathogen, because it results in a reported equine incidence of over 28,000 and mortality in over 300 bird species (Komar et al., 2003), causing substantial population declines among approximately 23 bird species (George et al., 2015; LaDeau et al., 2007). For example, the American crow (*Corvus brachyrhynchos*) population declined by 45% after the introduction of the WNV (LaDeau et al., 2007). The impact of WNV has not been limited to the US region, because more than 5,000 human infections have been reported in Canada, and the virus is recognized as an emerging threat across the Americas (Castro-Jorge et al., 2019; Hadfield et al., 2019).

In its natural cycle, WNV is maintained in birds and mosquitoes, although several wild vertebrates (Lichtensteiger et al., 2003; Osorio et al., 2012) and domestic animals such as horses, cats, and dogs can be naturally infected (Komar et al., 2003). Although WNV has been detected in 65 different mosquito species and 326 bird species only in the United States, only a few *Culex* mosquito species, such as *Culex pipiens* and *Culex quinquefasciatus*, drive the transmission of the virus in nature and the subsequent spread to humans (Hayes et al., 2005). Several avian species shed a significant number of viruses in oral or fecal secretions when infected, allowing direct bird-to-bird and even bird-to-human propagation (Mclean et al., 2001; Komar et al., 2003). This confirms the relevance of migratory birds in WNV transmission between different areas (Costa et al., 2019; Vilibic-Cavlek et al., 2019).

The development and expansion of WNV is extremely heterogeneous in each season, with periodic outbreak years combined with low levels of viral transmission, and climate is one of the key elements driving these scenario (Chung et al., 2013; Ruiz et al., 2010). For example, higher air temperatures influence vector population growth rates and its competence, accelerate virus replication in mosquitoes (extrinsic incubation period), prolong the breeding season (Costa et al., 2019) and increasing the efficiency of viral transmission to birds (Vilibic-Cavlek et al., 2019). Changes in climatic conditions have been hypothesized to play a central role in increasing the number of WNV outbreaks observed worldwide in the last decades. Viral transmission occurs when competent vectors are abundant and active (Chapman et al., 2018). In addition, precipitation is another climatic factor that have important consequences on mosquito productivity and abundance (Chuang et al., 2011; Degroote et al., 2014; Poh et al., 2019), thus precipitation also influences WNV transmission.

The recent Intergovernmental Panel on Climate Change report (IPCC, 2014) revealed an overall increase in warm days and heat waves in South America, and several other regions with more precipitation increases than decreases were observed, with heterogeneous spatial trends. Up to now, studies about the impact of climate change on WNV spreading in South America are limited. Although large-scale, the dissemination of this virus has not been accompanied by noticeable bird mortality or disease in horses or humans in Latin America and the Caribbean (Kramer et al., 2008); the main concern is precisely the absence of data on the disease burden in its hosts (Komar & Clark, 2006). Despite the availability of a comprehensive record in the literature for WNV, to the best of our knowledge, no predictive models have been developed in this context, especially for South America. In this study, we analyzed and illustrated how WNV can increase its relevance in the future. We addressed these questions by searching for evidence of virus circulation in human patients, mosquitoes, equids, and birds through South America during the last 15 years and developing a predictive model considering environmental variables that directly affect WNV transmission dynamics and distribution.

## 2. Materials and methods

### *2.1 Data acquisition*

The approximate locations of WNV records were determined using sites identified in the literature between 2004 (first record in South America) and 2020 (Table 1). Data were exhaustively collected using searches of the PubMed and Google Scholar databases. We included all records of WNV in South American regions reported in epidemiological bulletins since the very first record up until 2020. The criterion for inclusion of a region in the analysis was the presence of ≥ 1 recorded WNV.

The procedures followed in our study were based on Lorenz et al (2017). To determine the ecological and climatic conditions associated with the presence of WNV, we examined the associations between the locations of the records and eight variables: annual rainfall (mm), annual temperature (°C), elevation (m), type of biome, seasonality of temperature, seasonality of precipitation, thermal amplitude, and daytime temperature variation. The seasonality of the temperature value was calculated as the standard deviation of the average monthly temperature. The thermal amplitude value was calculated by subtracting the minimum temperature during the coldest month from the maximum temperature during the hottest month. The seasonality of the precipitation value was calculated as the coefficient of variation for the average monthly precipitation. The mean daytime temperature variation was calculated by subtracting the mean minimum temperature from the mean maximum temperature. All weather data were obtained in ASCII raster file format using the "LAT/LONG" geodetic coordinate system (Datum WGS-84). These data were obtained from the WorldClim - Global Climate Data database, which contains representative observational data for 1950±2000 that were interpolated to a resolution of 30 arc seconds (approximately 1 km). As the environmental variables were expressed in various units, principal component analysis (PCA) was performed after standardizing the variables using a Pearson correlation matrix.

**Table 1.** South American areas that have presented WNV detection in the 2004-2020 period (Data from Google Scholar and PubMed databases).

| COUNTRY | REGION | YEAR | RECORD TYPE | REFERENCE | COORDS |
|---|---|---|---|---|---|
| COLOMBIA | Córdoba | 2004 | Horse | Mattar et al., 2005 | 9.195296, -76.010268 |
| COLOMBIA | Sucre | 2004 | Horse | Mattar et al., 2005 | 8.844197, -74.745573 |
| ARGENTINA | Córdoba | 2005 | Bird | Diaz et al., 2008 | -31.760224, -63.560619 |
| COLOMBIA | El Meta | 2005 | Horse | Góez-Rivillas et al., 2010 | 3.828449, -73.898221 |
| ARGENTINA | San Antonio de Areco | 2006 | Horse | Morales et al., 2006 | -34.246975, -59.457666 |
| ARGENTINA | Victoria - Entre Rios Province | 2006 | Horse | Morales et al., 2006 | -32.624022, -60.150743 |
| ARGENTINA | Mar Chiquita - Córdoba | 2006 | Bird | Diaz et al., 2008 | -30.810762, -62.862172 |
| ARGENTINA | Chaco | 2006 | Bird | Diaz et al., 2008 | -26.162497, -60.722666 |
| ARGENTINA | Tucumán | 2006 | Bird | Diaz et al., 2008 | -26.819668, -65.216417 |
| ARGENTINA | Monte Cristo - Córdoba | 2006 | Bird | Diaz et al., 2008 | -31.339957, -63.945559 |
| ARGENTINA | Monte Alto | 2006 | Bird | Diaz et al., 2008 | -26.955432, -62.814174 |
| COLOMBIA | Atlántico - coast | 2006 | Horse | Mattar et al., 2011 | 10.912077, -74.991247 |
| COLOMBIA | Bolívar - coast | 2006 | Horse | Mattar et al., 2011 | 10.313668, -75.300519 |
| COLOMBIA | Magdalena - coast | 2006 | Horse | Mattar et al., 2011 | 11.149139, -73.982325 |
| VENEZUELA | Laguna de los Patos - Sucre | 2007 | Mosquito | Velásquez et al., 2013 | 10.423033, -64.201746 |
| COLOMBIA | Medelin | 2008 | Bird | Osorio et al., 2012 | 6.222606, -75.580183 |
| BRAZIL | Nhecolândia/MS | 2009 | Horse | Pauvolid-Corrêa et al., 2011; 2014 | -18.33861111, -57.24861111 |
| BRAZIL | Aquidauana/MS | 2009 | Horse | Raymondi-Silva et al., 2013 | -20.457118, -55.777345 |
| BRAZIL | Bonito/MS | 2009 | Horse | Raymondi-Silva et al., 2013 | -21.120316, -56.487325 |
| BRAZIL | Anastácio/MS | 2009 | Horse | Raymondi-Silva et al., 2013 | -20.487545, -55.811486 |
| BRAZIL | Bodoquena/MS | 2009 | Horse | Raymondi-Silva et al., 2013 | -20.551787, -56.678702 |
| BRAZIL | São José da Cruz do Brejo/PB | 2009 | Horse | Raymondi-Silva et al., 2013 | -6.229616, -37.342943 |
| BRAZIL | Nova Brasilândia/MT | 2009 | Horse | Ometto et al., 2013 | -14.773138, -55.060898 |
| BRAZIL | Lagoa do Peixe National Park/RS | 2009 | Bird | Ometto et al., 2013; Castro-Jorge et al., 2019 | -31.253750, -50.969954 |
| BRAZIL | Ilha da Canela/PA | 2009 | Bird | Ometto et al., 2013; Castro-Jorge et al., 2019 | -0.784771, -46.723208 |
| BRAZIL | Nabileque/MS | 2010 | Horse | Pauvolid-Corrêa et al., 2014 | -20.863440, -57.836708 |
| BRAZIL | Porto Jofre/MT | 2010 | Horse / chicken | Melandri et al., 2012 | -16.248888, -56.858888 |
| BRAZIL | Poconé/MT | 2010 | Horse / chicken | Melandri et al., 2012 | -16.665, -56.7938888 |
| BRAZIL | Pinheiro/MA | 2010 | Bird | Ometto et al., 2013; Castro-Jorge et al., 2019 | -2.528208, -45.089733 |
| COLOMBIA | Córdoba | 2012 | Mosquito | López et al., 2015 | 9.358333, -75.976944 |
| BRAZIL | Aroeiras do Itaim/PI | 2014 | Human | SESAPI (Secretaria de Saúde do Piauí) | -7.245577, -41.582601 |
| BRAZIL | Picos/PI | 2017 | Human | SESAPI (Secretaria de Saúde do Piauí) | -7.066997, -41.427460 |
| BRAZIL | Piripiri/PI | 2017 | Human | SESAPI (Secretaria de Saúde do Piauí) | -4.275203, -41.771882 |
| BRAZIL | São Mateus/ES | 2018 | Horse | Martins et al., 2019 | -18.721112, -39.847865 |
| BRAZIL | Baixo Guandu/ES | 2018 | Donkey | Silva et al., 2018 | -18.6883713, -40.4082043 |
| BRAZIL | Nova Venécia/ES | 2018 | Horse | Silva et al., 2018 | -18.702494, -40.329336 |
| BRAZIL | Lagoa Alegre/PI | 2019 | Human | SESAPI (Secretaria de Saúde do Piauí) | -4.513817, -42.621034 |
| BRAZIL | Teresina/PI | 2019 | Human | SESAPI (Secretaria de Saúde do Piauí) | -5.044479, -42.754136 |
| BRAZIL | Amarante/PI | 2019 | Human | SESAPI (Secretaria de Saúde do Piauí) | -6.238821, -42.840200 |
| BRAZIL | Água Branca/PI | 2020 | Human | SESAPI (Secretaria de Saúde do Piauí) | -5.905901, -42.628846 |
| VENEZUELA | Barinas | 2004-2006 | Horse / birds | Bosch et al., 2007 | 8.691259, -70.183141 |
| VENEZUELA | Zulia | 2004-2006 | Horse / birds | Bosch et al., 2007 | 10.481874, -71.794726 |
| VENEZUELA | Zulia | 2004-2006 | Horse | Bosch et al., 2007 | 10.589885, -72.256152 |
| VENEZUELA | Zulia | 2004-2006 | Horse | Bosch et al., 2007 | 8.944228, -71.618945 |
| VENEZUELA | Yaracuy | 2004-2006 | Horse | Bosch et al., 2007 | 10.495604, -68.827474 |
| VENEZUELA | Yaracuy | 2004-2006 | Horse | Bosch et al., 2007 | 10.006410, -68.563802 |
| VENEZUELA | Carabobo | 2004-2006 | Horse | Bosch et al., 2007 | 10.249961, -68.071230 |
| VENEZUELA | Guarico | 2004-2006 | Horse | Bosch et al., 2007 | 9.784117, -66.051638 |
| VENEZUELA | Guarico | 2004-2006 | Horse | Bosch et al., 2007 | 9.193557, -65.425417 |
| VENEZUELA | Guarico | 2004-2006 | Horse | Bosch et al., 2007 | 9.144750, -65.700075 |
| VENEZUELA | Guarico | 2004-2006 | Horse | Bosch et al., 2007 | 8.830059, -67.089846 |
| VENEZUELA | Guarico | 2004-2006 | Horse | Bosch et al., 2007 | 8.547693, -66.650393 |
| VENEZUELA | Anzoategui | 2004-2006 | Bird | Bosch et al., 2007 | 8.762900, -64.179341 |
| VENEZUELA | Sucre | 2004-2006 | Bird | Bosch et al., 2007 | 10.301206, -64.183431 |
| COLOMBIA | Turbo - Antioquia | 2006-2008 | Horse | Góez-Rivillas et al., 2010 | 8.094207, -76.731325 |
| COLOMBIA | Chigorodó - Antioquia | 2006-2008 | Horse | Góez-Rivillas et al., 2010 | 7.633846, -76.620891 |
| COLOMBIA | Bolombolo - Antioquia | 2006-2008 | Horse | Góez-Rivillas et al., 2010 | 5.970876, -75.838147 |

*2.2 Data analysis*

PCA was performed using XLSTAT software (Addinsoft, 2020) to preselect the environmental variables that were used in the WNV modeling. The PCA approach was used for two reasons. First, PCA facilitates the identification and elimination of covariant variables, which is a key procedure for avoiding analytical artifacts. Second, PCA has been widely used in equivalent studies and facilitates comparisons, reproducibility, and future meta-analysis. After performing the PCA, we selected the four most representative eigenvectors of the variables, which were used for Maxent analysis (version 3.3.3 k: a machine learning algorithm for modeling species distributions based on existing data and environmental variables) (Phillips & Dudík, 2008; Elith et al., 2011). Data selection was performed according to the criterion of maximum entropy, with the original variables that reached maximum and minimum values within the ordered ranking of each principal component, because they describe the full range of data variation. The Maxent model may be expressed as:

$$\mathbf{p}[\mathbf{f_j}] = \frac{\mathbf{1}}{\mathbf{N}}\sum_{\mathbf{i=1}}^{\mathbf{N}} \mathbf{f_i}(\mathbf{x_i})$$

Where: $x^*$ = the geographical region of interest; $x = \{x_1, x_2 \dots, x_N\}$ with $x \in x^*$; $x \rightarrow$ observed points at $x^*$; $f_j = f_{1'}, f_{2'} \dots, f_m$ (environmental variables); $N$ = the number of observed records; and $p$ = the probability of WNV occurrence. The model was run 25 times, with a difference of 10% of the localities for each run to estimate the parameters and their precision. Potential distribution maps were created by interpolating the occurrence points and the similarity measures of the environmental variables in each pixel (i.e., a known observation probability value can be assigned to each pixel by calculating a probability whose exponent is a quadratic function).

Future climate data were integrated using the Model for Interdisciplinary Research on Climate (MIROC5) (Watanabe et al., 2011) global climate model (GCM). The MIROC5 model includes components of the Earth's system and climate change in relation to anthropogenic radiation. The advantage of using this model is that it increases the accuracy of short-term climate prediction, as it can be affected by both anthropogenic and intrinsic fluctuations of the climate system. The spatial resolution of the GCM was the same as that of the environmental variables (30 arc seconds, approximately 1 km). The comparison method was the same as for the Maxent model, although the probability calculation for the GCMs incorporated a comparison of the present and future environmental conditions. To obtain future climate scenarios using GCMs, it is also necessary to choose a condition for the evolution of greenhouse gas emissions (GGEs) during the period when the future climate is projected. In our prediction, we used two different scenarios: low emissions (Representative Concentration Pathway [RCP] 2.6) and very high emissions (RCP 8.5), detailed in the Special Report on Emissions Scenarios by the Intergovernmental Panel on Climate Change (Pachauri et al., 2015). In the first case, the global temperature tends to increase by 1.0ºC and can reach a temperature anomaly ranging from 0.4 to 1.6ºC and 0.3 to 1.7ºC between 2046–2065 and 2081–2100, respectively (Van Vuuren et al., 2007). In the second scenario, with high GGE, the global temperature tends to increase 2.0 to 3.7ºC and can reach to a thermal anomaly ranging from 1.4 to 2.6ºC and 2.6 to 4.8ºC between 2046–2065 and 2081–2100, respectively (Chou et al., 2014; Riahi et al., 2007).

The models of future expansion of WNV in South America were subsequently projected into the timeline and the two future climatic conditions (2046–2065 and 2071–2100) to identify areas suitable for the virus circulation. The default Maxent auto-feature setting was used (linear, quadratic, product, threshold, and hinge). The maps were edited using QGis software 2.10.1.

## 3. Results

Regarding PCA, the first two components (F1 and F2) accounted for 66.04% of the variation (Figure 1). According to this analysis, it was possible to perceive that the most representative variables of PC were associated with precipitation (annual rainfall and precipitation seasonality) and temperature (annual temperature average and thermal amplitude).

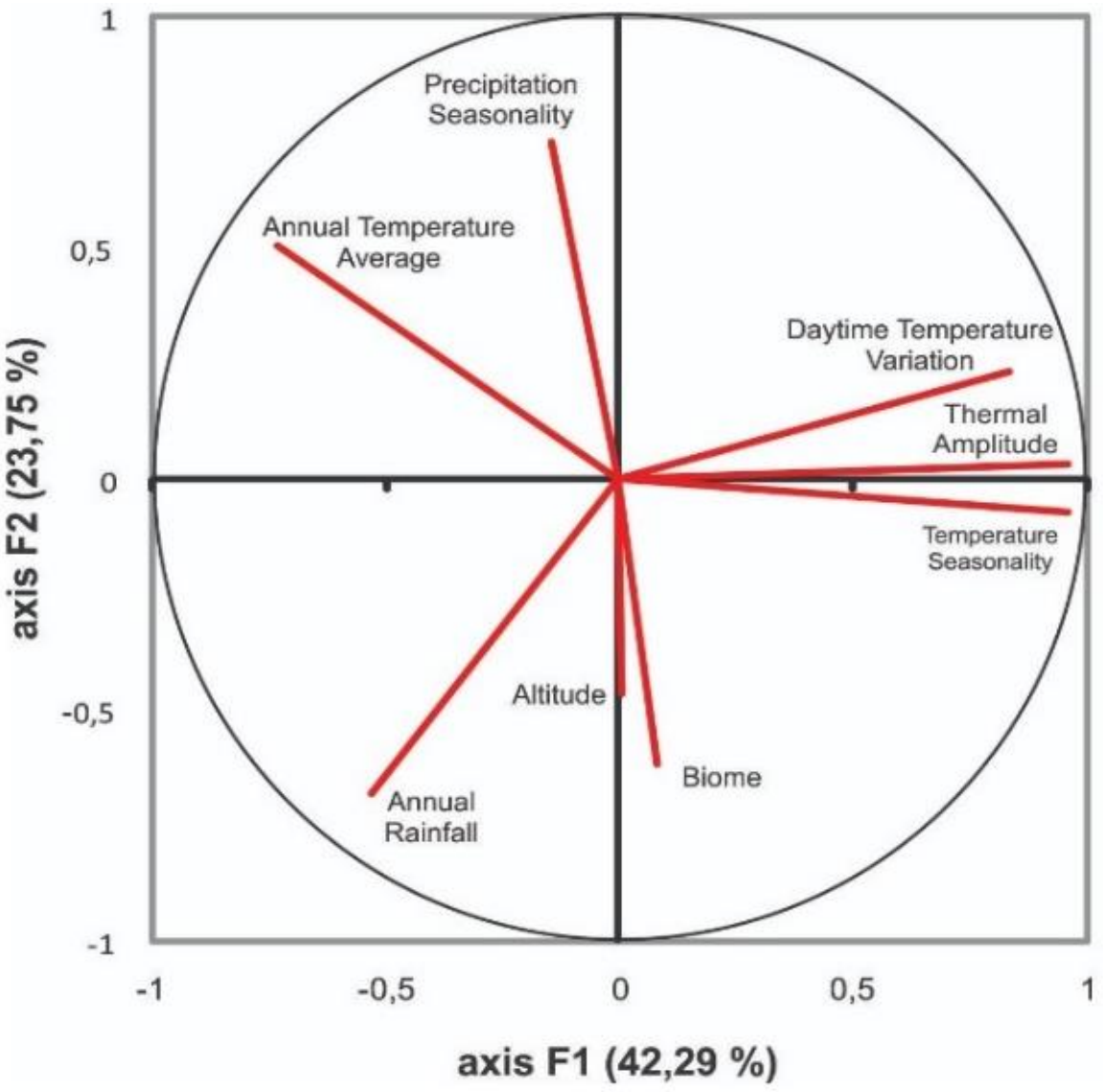

| Variables | Eigenvectors | |
|---|---|---|
| | F1 | F2 |
| Biome | 0,043 | -0,447 |
| Altitude | 0,003 | -0,335 |
| Annual Rainfall | -0,287 | **-0,492** |
| Daytime Temperature Variation | 0,454 | 0,171 |
| Annual Temperature Average | **-0,399** | 0,366 |
| Precipitation Seasonality | -0,078 | **0,529** |
| Thermal Amplitude | **0,523** | 0,027 |
| Temperature Seasonality | 0,520 | -0,050 |

**Figure 1.** Principal Component Analysis. The most representative variables are highlighted.

After selecting the four most important variables, we constructed a Maxent model (Figure 2) to determine the areas were WNV will most likely to be observed in South America. We observed that there is a concentration of WNV in the northern region of South America, including Venezuela and Colombia, eastern Brazil, and mainly countries such as Paraguay, Bolivia, and South Argentina. The contribution of each variable to the final model in Maxent was annual rainfall of 51.6%, annual temperature of 36.3%, precipitation seasonality of 6.5%, and thermal amplitude of 5.5%. This final model had an area under the curve of 0.85, which was significantly better than the random prediction (p = 0.001), indicating good performance of the model. The behavior of each variable in the model is shown in Figure 3.

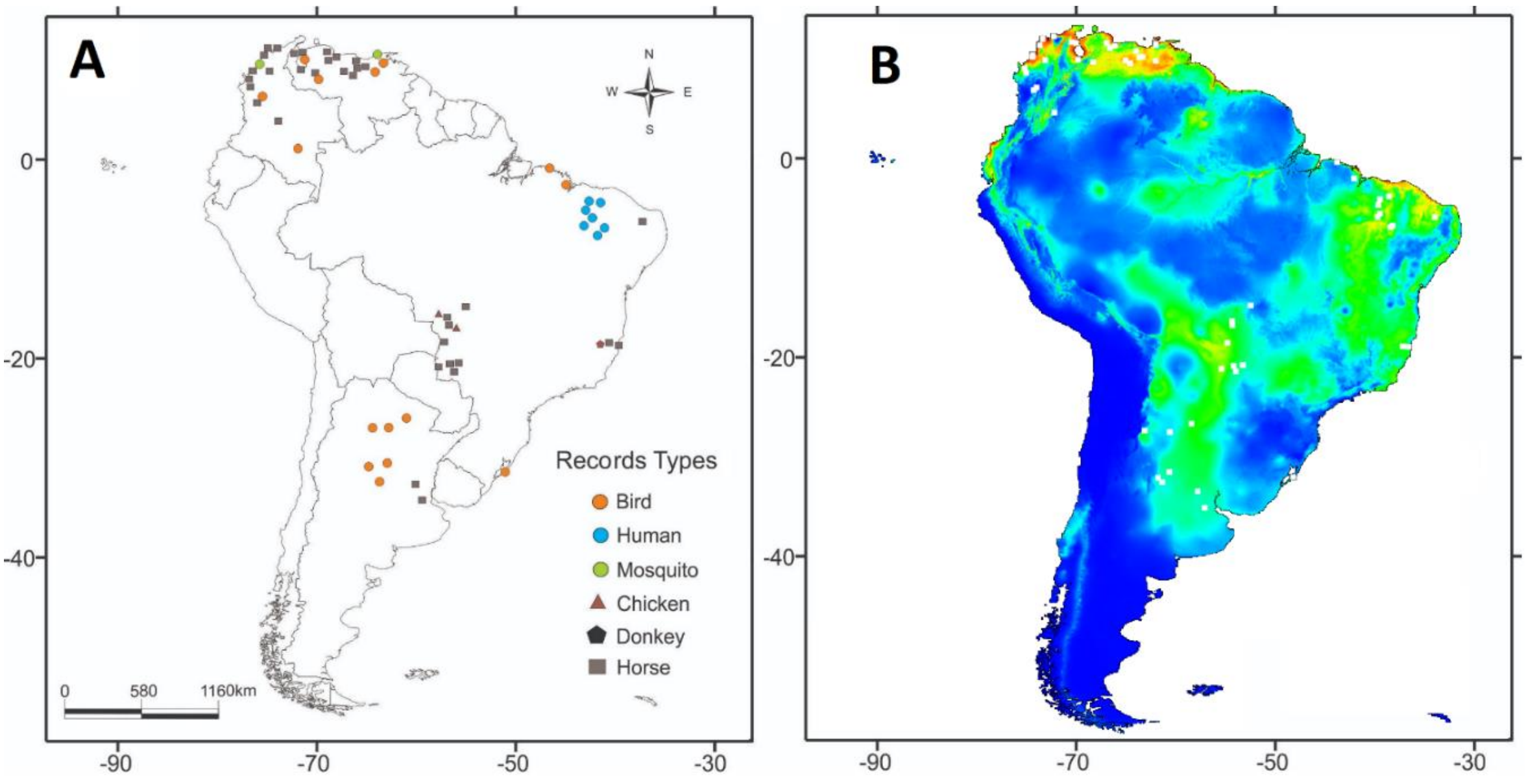

**Figure 2.** Occurrence of WNV in South America during the last 15 years. **A**. Records by host type. **B**. Maxent model showing the probable current distribution. The maps were built using QGis software 2.10.1.

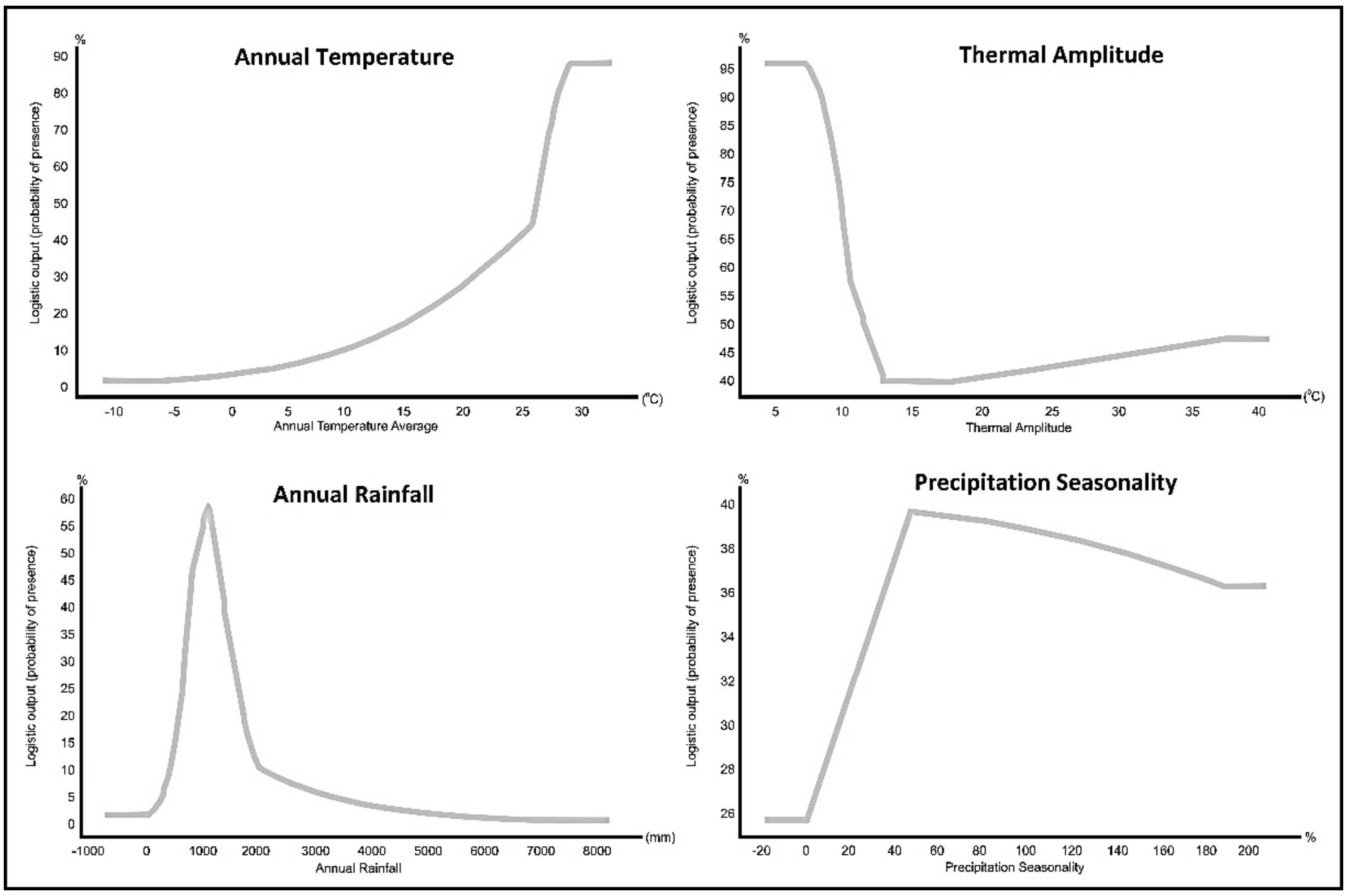

**Figure 3.** Behavior of each selected variable in the Maxent model according to the logistic output (probability of presence of WNV).

Probability maps were generated using two different future climate projections (Figure 4). The results reveal progressively expanding areas with an increased likelihood of WNV distribution, especially in areas with high GGE levels. Countries such as Bolivia and Paraguay will be greatly affected, drastically changing their current WNV distribution. Several Brazilian areas will also increase the likelihood of presenting WNV, mainly in the Northeast and Midwest regions. It is worth highlighting the possible change in the distribution of WNV on the Galápagos Islands in the following years (Figure 5).

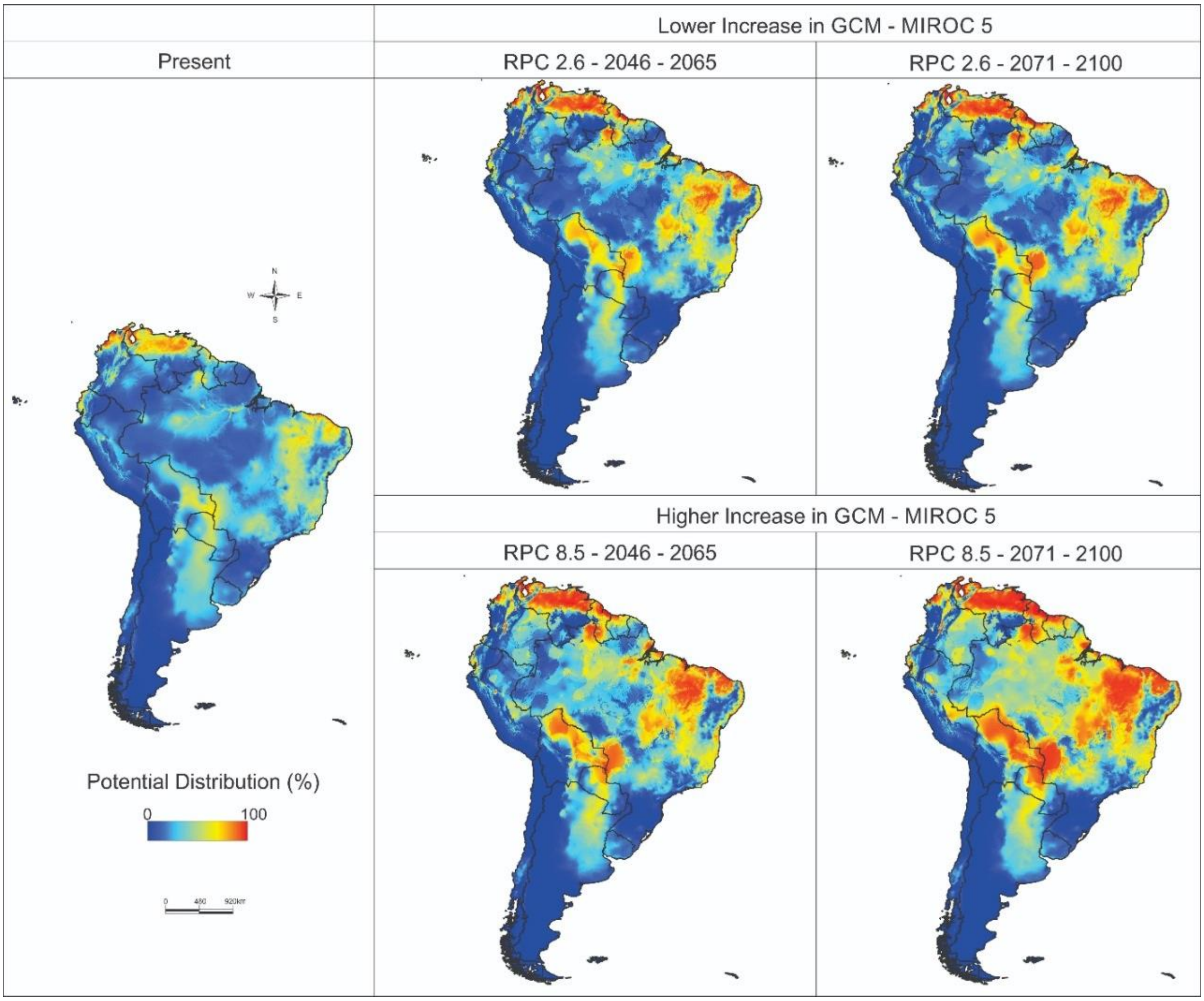

**Figure 4.** Predicted WNV range expansion in South America based on GCM MIROC-5. The maps show the distribution under two climate change scenarios: RCP 2.6 (lower increase in greenhouse gas emissions) and RCP 8.5 (higher increase in greenhouse gas emissions). The maps were built using QGis software 2.10.1.

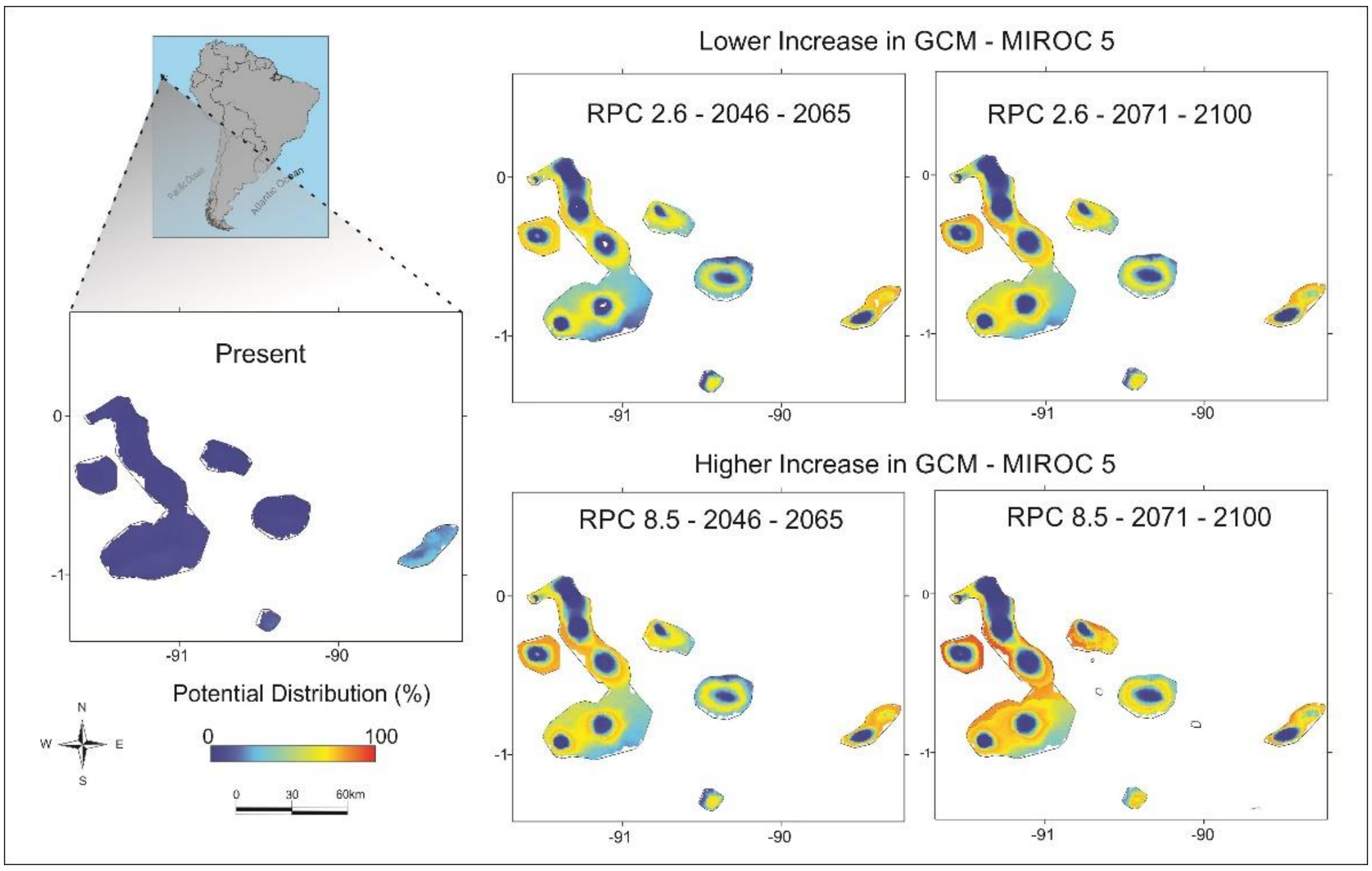

**Figure 5.** Current and predictive distribution of WNV on the Galapagos Islands.

## 4. Discussion

Predicting at-risk locations for the WNV is important for targeting vector control and public health resources. According to the reviewed literature, there are several records of WNV distribution in South America detected in different types of hosts, but our Maxent model showed that the distribution area of this virus can be significantly larger than observed. It is possible that there may be a significantly larger number of infected mosquitoes, horses, and birds, but the lack of adequate testing and serological surveys among populations may underestimate this number. The incongruence between the reported cases and the real cases is mainly due to the similarity of the clinical symptoms among many arboviruses, underestimating the occurrence of WNV in South America. This scenario is aggravated by the lack of accurate diagnostic methods for identifying the virus. Considering that several countries in South America have been facing a major dengue, Zika, and chikungunya outbreaks in last decades (WHO, 2020), it is relatively possible that other arboviruses with related symptoms have been confused with WN fever itself. Although the knowledge of the high circulation of WNV in some areas, several cases remain undiagnosed, probably because of their usually mild and self-limited clinical manifestations. Patients generally recover completely after a few days. However, more severe cases may remain undiagnosed, especially because of the long distances of traveling to healthcare facilities, transport difficulties of the sample, and lack of laboratories capable of conducting diagnostic tests (Lorenz et al., 2017).

Perhaps the WNV dissemination in South America behaves differently from that observed in North America and Europe, because the high variety of viruses may increases the level of genetic resistance in South America. WNV can circulate in the same manner as other flaviviruses (Bosch et al., 2007; Pauvolid-Corrêa et al., 2011). These other flaviviruses infect birds and equids but do not cause severe clinical symptoms that are important to health surveillance. Apparently, South American countries appear to have established a form of coexistence with WNV in which the virus can circulate, while at a low level in terms of pathogenicity, without causing major problems (Ometto et al., 2013). Since the introduction of WNV in South America in 2004, only seven human cases have been recorded in Brazil, with only one death reported (MS, 2020).

Our results confirm the associations between the WNV modeling literature and temperature and precipitation in the eastern United States, with higher temperatures and lower precipitation associated with increased WNV incidence (Hahn et al., 2015). Higher temperatures result in greater fecundity, faster time-to-development, and increased blood meal intake in *Culex* mosquitoes that transmit WNV (Ciota et al., 2014). Increased adult mosquito populations and more aggressive feeding behavior on reservoir species both increase the probability of detecting WNV at a trap site. Precipitation amount had an inverse association with WNV incidence, confirming the result of a study conducted in Illinois (Ruiz et al., 2010). *Culex* larvae and pupae are particularly susceptible to being flushed out of the containers they inhabit and are easily killed by rainfall events (Gardner et al., 2012; Koenraadt and Harrington, 2008; Myer & Johnston, 2019). Heavy rainfall might also dilute the nutrients for larvae, thus decreasing their development rate (Chevalier et al., 2013). We found that approximately 1200 mm of rainfall was the optimal value for maximizing the potential distribution of the WNV. A possible explanation is that drought events lead to close contact between avian hosts and mosquitoes around the remaining water sources, thereby accelerating the epizootic cycling and amplification of WNV within these populations (Shaman et al., 2005). Furthermore, during drought conditions, standing water pools are richer in organic materials that mosquitoes need to thrive (Paz et al., 2013). Such water areas might also be attractive for several bird species, which might favoring the bird–mosquito interaction (Paz, 2015).

According to the predictive MIROC-5 model, high-risk areas for WNV may change over the next few decades. This scenario can be more or less dramatic and depends on GGE levels. Countries such as Bolivia and Paraguay will be greatly affected, drastically changing their current WNV distribution. Several Brazilian areas will also increase the likelihood of presenting the WNV, mainly in the Northeast and Midwest regions. Climatic change also affects human activity, human migration, and vector redistribution, resulting in a more favorable environment for the propagation of arboviruses (Pignatti, 2004). Therefore, future risk estimations should consider these factors. However, despite all efforts and several studies performed, it remains challenging to identify the main cause of an outbreak (Massad, 2008) and to define the most efficient method(s) for protecting humans from viruses. The WNV cycle is complex and its details vary by region, making it difficult to model on a broad scale (Kramer et al., 2008), and we showed that temperature and precipitation-related variables appear to play a fundamental role in the epidemiology of WNV.

Our findings also showed that the Brazilian Pantanal biome greatly increases the current distribution of WNV. This finding is possibly associated with the specific environmental and ecological characteristics of this area, such as the predominance of migratory birds from North America (Nunes & Tomas, 2008). This region is also characterized by strong anthropogenic disturbances, including the recent expansion of farming, which involves human activities linked to deforestation. These factors can interfere with the natural cycle of the virus and increase the risk of WNV transmission to humans (Ometto et al., 2013). In contrary, looking at the current and future distribution maps, it is possible to see that the Andes Mountains act as a physical barrier against the dispersion of the WNV in South America, which can be related to the absence of bird migration paths in this area.

With the expansion of the geographical range of the WNV, it is almost certain that more epidemics of meningoencephalitis will be reported in the predictable future. Cities with poor economic and infrastructure conditions and those that lack effective arbovirus surveillance systems and mosquito vector control programs are particularly vulnerable (Campbell et al., 2002). Considering the large population of reservoir birds and the abundance and diversity of mosquito species in South America, WNV may have become endemic in some regions. Among the several species in the genus *Culex* reported in South America (Tissot & Silva, 2008), *Cx. quinquefasciatus* is the most abundant anthropophilic species (Consoli & Lourenço-de-Oliveira, 1994). Some studies point out that the emergence and dissemination of *Culex*-transmitted diseases probable will be intensified in the coming years due to their strong adaptation to climate change and ineffective urban sanitary infrastructure. The continuous evolution of WNV in North America (McMullen et al., 2011) and the potential effects of WNV dissemination in South America on economy and public health highlight the urgency to establish WNV active surveillance and research programs. It’s necessary to minimize possible damage to society from the introduction of another lethal arbovirus in addition to those already established, such as dengue, Zika, chikungunya, and yellow fever (Ometto et al., 2013).

Our study used several records of migratory birds from different stopover points, and according to our results, its possible that the entrance of WNV in South America is most likely due to birds resting at suitable “staging posts” on their north-to-south travel. Presumably these birds have progressively moved the virus to the south in stages, rather than clinically infected birds traveling the full distance. These data have been supported by the isolation of WNV antibodies in equids only since 2009, along with other studies that have also reported positive serology during the same period (Melandri et al., 2012; Pauvolid-Corrêa et al., 2011). The introduction of WNV in Argentina occurred early through birds in January 2005 (Diaz et al., 2008). This information reinforces the concept of staging posts, as there are bird displacements between Brazil and Argentina, and relatively high contact occurs among these posts (Ometto et al., 2013). It is important to mention that beyond continental migration routes, there are several regional migration routes that can help to further spread the WNV. Among the existing regional migratory routes in South America, one of the most important is the Central Depression route in Rio Grande do Sul State, which is located along the coast of the Atlantic strip of Uruguay to the south of Santa Catarina State in Brazil. To reach Argentina, birds use the natural corridor of rivers, small lagoons, and wetlands of the Central Depression. Using this route, a confirmed case of WNV in birds has been reported.

We found that the Galápagos Islands will probably increase their geographic range suitable for WNV occurrence. The wildlife species of the Galápagos Islands have particular characteristics that evolved in isolation over thousands of years, with high levels of endemism in the fauna and flora (Eastwood et al., 2014). Nevertheless, these unique ecosystems of island are experience modern pressures from an increasing number of human population, tourism, invasive species, and disease introduction (Cruz et al., 2009). WNV is related with population decreases in many North American bird species (LaDeau et al., 2007). Thus, this virus represents a relevant threat to the native species of Galápagos, when WNV reaches the archipelago (Peterson et al., 2004). The most likely pathway for a WNV introduction event to Galápagos is predicted to be human-mediated, chiefly by the transport of infected mosquitoes to the islands by airplane (Kilpatrick et al., 2006). Invasive insects are known to enter Galápagos via this route (approximately one-fourth of all insect fauna in Galápagos are foreign (Causton et al., 2006). A recent study detected *Cx. quinquefasciatus* mosquitoes inside an aircraft arriving at Galápagos (Baltra Airport), and genetic analysis has indicated that there have been multiple introductions of this mosquito, although pathogen infection status was not ascertained (Bataille et al., 2009; Eastwood et al., 2011).

This study has some limitations. First, the projections presented here were processed based on the assumption that all variables remained stable over time, except the temperature. For example, the human population size was considered stable in our models. Second, we did not take into account the deforestation rate, which are important factor in determining WNV outbreaks. However, other factors that were not part of our model could change over the given time period, such as the quality of vector surveillance, clinical case detection, and the development of some vaccines. In this study we did not considered the distribution areas of the mosquito vectors, which are essential for transmission, but are significantly difficult to estimate. Finally, WNV may circulate in asymptomatic or misdiagnosed patients, and it is not possible to determine their exact distribution, which directly affects the predictive power of our model. Despite the number of limitations in our present study, it is an important initial step in predicting the emergence of WNV in South America. It also serves to warn the vigilant surveillance healthcare systems that WNV may remain underestimated before the outbreak occurs.

## 5. Conclusions

This is the first spatiotemporal study in South America to predict WNV distribution. Our findings showed that the probable high-risk areas could be significantly larger than those in which the WNV was detected. Environmental factors can directly affect the distribution of WNV, with higher temperatures and lower precipitation associated with increased virus incidence. High-risk areas may be modified in the coming years, being more pronounced in areas with high GGE levels. Countries such as Bolivia and Paraguay will be greatly affected, drastically changing their current WNV distribution. Several Brazilian areas will also increase the likelihood of presenting the WNV, mainly in the Northeast and Midwest regions and the Pantanal biome. The Galápagos Islands will also probably increase their geographic range suitable for WNV occurrence.

Although our approach is limited, we propose that future studies should be multidisciplinary and include other variables such as vector distribution and components of the natural cycle of WNV. Overall, this study provides useful indications regarding the dynamics of WNV in the South American region. Considering the reservoirs of migratory birds and *Culex* mosquitoes present in South America and the potential availability of new animal reservoirs due to the vast size of this region, it is only a matter of time before the WNV spreads across the countries once its replication cycle has been established in the environment, as was observed following its emergence in the United States.

It is required to increase active epidemiological surveillance in animals and humans and promote preventive actions to minimize the possibility of WNV infection in humans before it becomes a major public health issue, as observed with other arbovirus infections. WNV will also likely continue to spread into Central and South America, but the public health consequences of this dissemination remain uncertain.

**Funding**

This work was supported by São Paulo Research Foundation (FAPESP) grant number 2017/10297-1.